\documentclass[
    superscriptaddress,
    reprint,
    showkeys,
    showpacs,
    amsmath,
    amssymb,
    aps,
    prb,
    floatfix
]{revtex4-1}
\usepackage{physics}
\usepackage{siunitx}
\usepackage{graphicx}
\usepackage{tabularx}
\usepackage{float}
\usepackage[caption=false]{subfig}

\begin{document}

\title{Field-effect induced superconductivity in surfaces of tetrahedrally coordinated semiconductors: the case of (111) hydrogenated silicon}
\author{D. Romanin}
\email{davide.romanin@polito.it}
\affiliation{
    Department of Applied Science and Technology, Politecnico di Torino, 10129 Torino, Italy
}
\date{\today}

\begin{abstract}
We show the possibility of inducing a superconductive phase transition in tetrahedrally coordinated semiconductors via field-effect (FET) doping by taking as an example the hydrogenated (111) silicon surface. We perform density functional theory computations of the electronic and vibrational properties of the system in the proper FET geometry, by taking into account the applied electric field and the induced charge density. Using a simplified superconductive model at $\vb{q}=\vb{\Gamma}$ and the McMillan/Allen-Dynes formula, we get an estimate of the superconductive critical temperature. We observe that, by heavily doping with holes at $n_{dop}=6\cdot10^{14}$ cm$^{-2}$, we get an electron-phonon coupling constant of $\lambda_{Si}=0.98$ and a superconductive phase transition at $T_{\text{c}}\in[8.94;10.91]$ K, with $\mu^*\in[0.08;0.12]$.
\end{abstract}

\maketitle

\section{Introduction}

Since the pioneering work of M. L. Cohen in 1964  \cite{ref:CohenPR1964}, degenerate semiconductors have been shown both theoretically \cite{ref:ConnetablePRL2003, ref:BoeriPRL2007, ref:BlaseNM2009} and experimentally \cite{ref:EkimovNat2004, ref:BustarretNat2006} to undergo a superconductive phase transition upon chemical doping. In 2004 Ekimov et al. \cite{ref:EkimovNat2004} observed a superconductive transition with critical temperature $T_c\sim4$ K in boron-doped bulk diamond (with a dopant concentration $n_B\approx4\times10^{21}$ cm$^{-3}$), while in 2006 Bustarret et al. \cite{ref:BustarretNat2006} showed that the same happens for boron-doped bulk silicon, at $T_c\sim0.35$ K ($n_B\approx3\times10^{21}$ cm$^{-3}$). Even if ab-initio computations \cite{ref:ConnetablePRL2003, ref:BoeriPRL2007, ref:BlaseNM2009} suggest that the critical temperature could be enhanced by incrementing the amount of dopants, the solubility limit hinders further inclusion of boron into the crystal structure.

A possible alternative to chemical doping is field-effect doping \cite{ref:DagheroPRL2012, ref:PiattiPRB2017}, which induces an accumulation of charges in the first few layers of the sample. The technique consists in applying an electric field between the sample and a gate electrode, in a field-effect-transitor (FET)  configuration. If a solid dielectric is used to separate the sample from the gate, surface densities of induced charges as high as $10^{12}-10^{13}$ cm$^{-2}$ can be obtained. Much higher values (up to $10^{14}-10^{15}$ cm$^{-2}$ \cite{ref:DagheroPRL2012, ref:PiattiPRB2017})  can be attained by replacing the solid dielectric with a polymer-electrolyte solution or an ionic liquid. When a positive (negative) gate voltage is applied, the anions (cations) present in the electrolyte accumulate at the interface with the sample, inducing in its first few layers a negative (positive) charge distribution.

Following the theoretical prediction of field-effect induced superconductivity in hole-doped hydrogenated diamond surfaces \cite{ref:PiattiLTP2019, ref:NakamuraPRB2013, ref:SanoPRB2017, ref:RomaninAPSUSC2019}, in this work we perform a similar analysis using ab-initio density functional theory on the hydrogenated silicon surface. More precisely, since high-$T_c$ superconductivity was found in the (111) hydrogenated diamond surface (H-C(111)) at high doping \cite{ref:RomaninAPSUSC2019}, we will here investigate the (111) hydrogenated surface of silicon (H-Si(111)) when doped via field effect at the same hole concentration.

\section{Model and methods}
\subsection{Computational methods}

\begin{figure}
\centering
\includegraphics[width=0.6\linewidth]{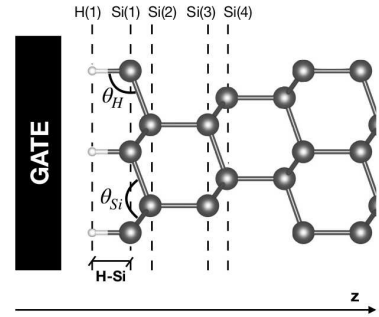}
\caption{Stick-and-ball model of the hydrogenated silicon (111) surface (side view). $\theta_H$ is the angle formed by H(1) and Si(1), while $\theta_{Si}$ is the angle formed by Si(1) and Si(2). $H-Si$ denotes the bond length between H(1) and Si(1).} \label{fig:atomic_structure}
\end{figure}
In this work we model the H-Si(111) surface as a slab made of 14 atomic layers, terminated on both sides by a layer of H atoms, for a total of 16 atoms in the primitive cell  (Fig.~\ref{fig:atomic_structure}). The slab will be centered around $z=0$ (where $z$ is the axis perpendicular to the surface) in order to have a symmetric system. Calling $L$ the total length of the cell used for computations, the layer of accumulated charges at the interface will be modelled as a planar uniform charge distribution placed at $z_{gate}=-0.181L$, while a potential barrier of height $V=6$ Ry is set at $z=-0.18L$. In order to avoid spurious interactions between repeated images of the system due to periodic boundary conditions, we add $\approx30$ \AA~of vacuum along $z$. The H-Si(111) surface is obtained by starting from bulk Si, whose lattice parameter was computed to be $a_{B,t}=5.46859$~\AA (which is only $0.7\%$ larger than the experimental value \cite{ref:Wyckoff1963} $a_{B,e}=5.43070$~\AA). In order to have a comparison with the induced high-$T_{c}$ superconductive phase of the H-C(111) surface \cite{ref:RomaninAPSUSC2019}, in this work we will investigate the same \textit{hole} doping: $n_{dop}=6\cdot10^{14}$ cm$^{-2}$.

Density functional theory (DFT) computations are performed through the Quantum ESPRESSO package \cite{ref:QE1,ref:QE2,ref:SohierPRB2017}. Hole doping and the presence of a transverse electric field due to the FET configuration are taken fully into account in a self-consistent way, both for electronic and vibrational properties, as described in Ref.~ \cite{ref:SohierPRB2017}, and the appropriate boundary conditions for ground state and linear response computations are set by truncating the long-range Coulomb interaction along the non-periodic $z$ direction. Exchange and correlation are here modelled according to the Perdew-Burke-Ernzerhof (PBE) functional, while the interaction between valence electrons and the core are taken into account through ultrasoft pseudopotentials \cite{ref:UFP} for both atomic species. The First Brillouin Zone (FBZ) is sampled, both for the neutral and doped surface, with a Monkhorst-Pack grid of $24\times24$ electron momenta ($\vb{k}$-points). Self-consistency is checked upon satisfaction of convergence criteria, i.e. $10^{-9}$ Ry for the total energy and $10^{-3}$ Ry/a$_0$ for the total force per atom. The cut-off for the kinetic energy is set to $30$ Ry, while that for the charge density is set to $240$ Ry.
We use a Gaussian smearing of $0.006$ Ry for the doped surface, and of $0.003$ Ry for electronic and vibrational computations respectively. The value of the smearing is chosen so as to ensure convergence of the total energy per atom ($<1$ mRy) and of the total force per atom ($<1$ mRy/a$_0$), with the additional requirement that it is smaller than the difference between the Fermi level and the top of the last crossed valence band. Finally, when computing the electronic density of states (DOS) we increment the $\vb{k}$-point uniform grid to $96\times96$.

\subsection{Simplified superconductive model}
\label{sec:model}
In order to estimate the superconductive critical temperature ($T_c$), we will employ the McMillan/Allen-Dynes formula~ \cite{ref:McMillanPR1968,ref:AllenPRB1975}:
\begin{subequations}
\begin{eqnarray}
T_{c} = &&\frac{\omega_{\textrm{log}}}{1.2}\exp\Bigl\{-\frac{1.04(1+\lambda)}{\lambda - \mu^*(1+0.62\lambda)}\Bigr\}\label{eq:AD}\\
\lambda = &&2\int d\omega \frac{\alpha^2F(\omega)}{\omega}\label{eq:lambda}\\
\omega_{\textrm{log}} = &&\exp\Bigl\{\frac{2}{\lambda}\int\log(\omega)\frac{\alpha^2F(\omega)}{\omega}d\omega\Bigr\}\label{eq:omega_log}
\end{eqnarray}
\end{subequations}
where $\lambda$ is the electron-phonon coupling constant 
and  $\omega_{\textrm{log}}$ is the logarithmic averaged phonon frequency. 
The Eliashberg spectral function $\alpha^2F(\omega)$ expresses the frequency dependence of the electron-phonon interaction and is defined by:
\begin{equation}
\label{eq:spectral}
\begin{split}
\alpha^2F(\omega)=&\frac{1}{N_{\sigma}(0)N_qN_k}\sum_{\nu,n,m}\sum_{\vb{k},\vb{q}}\abs{g^\nu_{\vb{k}n,\vb{k}+\vb{q}m}}^2\delta(\epsilon_{\vb{k}n})\\
&\delta(\epsilon_{\vb{k}+\vb{q}m})\delta(\omega-\omega_{\vb{q}\nu})
\end{split}
\end{equation}
where $N_{\sigma}(0)$ is the total electronic density of states per spin at the Fermi level (here we set $E_F=0$), $N_k$ and $N_q$ indicate the total number of electronic $\vb{k}$-points and phonon $\vb{q}$-points in the irreducible FBZ used for the sum. Finally,  $g_{\vb{k}n,\vb{k}+\vb{q}m}^{\nu}$ represents the electron-phonon matrix element between bands $n$ and $m$ for the phonon mode $\nu$:
\begin{equation}
\label{eq:elphon}
g_{\vb{k}n,\vb{k}+\vb{q}m}^{\nu} =\sum_{A\alpha} \frac{e^{A\alpha}_{\vb{q}\nu}}{\sqrt{2M_A\omega_{\vb{q}\nu}}}\bra{\vb{k}n}\fdv{v_{\text{SCF}}}{u_{A\alpha}^{\vb{q}}}\ket{\vb{k}+\vb{q}m}\,.
\end{equation}
In the equation above, $v_{\text{SCF}}=e^{-i\vb{q}\cdot\vb{r}}V_{\text{KS}}$ is the periodic part of the Kohn-Sham potential $V_{\text{KS}}$, $\ket{\vb{k}n}$ is the Bloch-periodic part of the Kohn-Sham eigenfuction, and $A$ labels atoms in the unit cell whose mass is denoted by $M_A$ and whose cartesian coordinates are indicated by $\alpha$. The phonon eigenvector normalized on the unit cell is here denoted by $e^{A\alpha}_{\vb{q}\nu}$, and the Fourier transformed displacement of atom $A$ along the cartesian direction $\alpha$ by $u_{A\alpha}^{\vb{q}}$. While $\lambda$ and $\omega_{\textrm{log}}$ can be computed ab-initio, the Morel-Anderson pseudopotential $\mu^*$ (i.e. the screened effective Coulomb interaction between electrons) is an ad-hoc parameter: we will assume that it takes values in the same range as in superconductive bulk silicon \cite{ref:JinJPCM1997, ref:BourgeoisAPL2007}, i.e. $\mu^{*}\in[0.08;0.12]$.

\begin{figure}[ht]
\centering
\subfloat[]
	{ \includegraphics[width=0.5\linewidth]{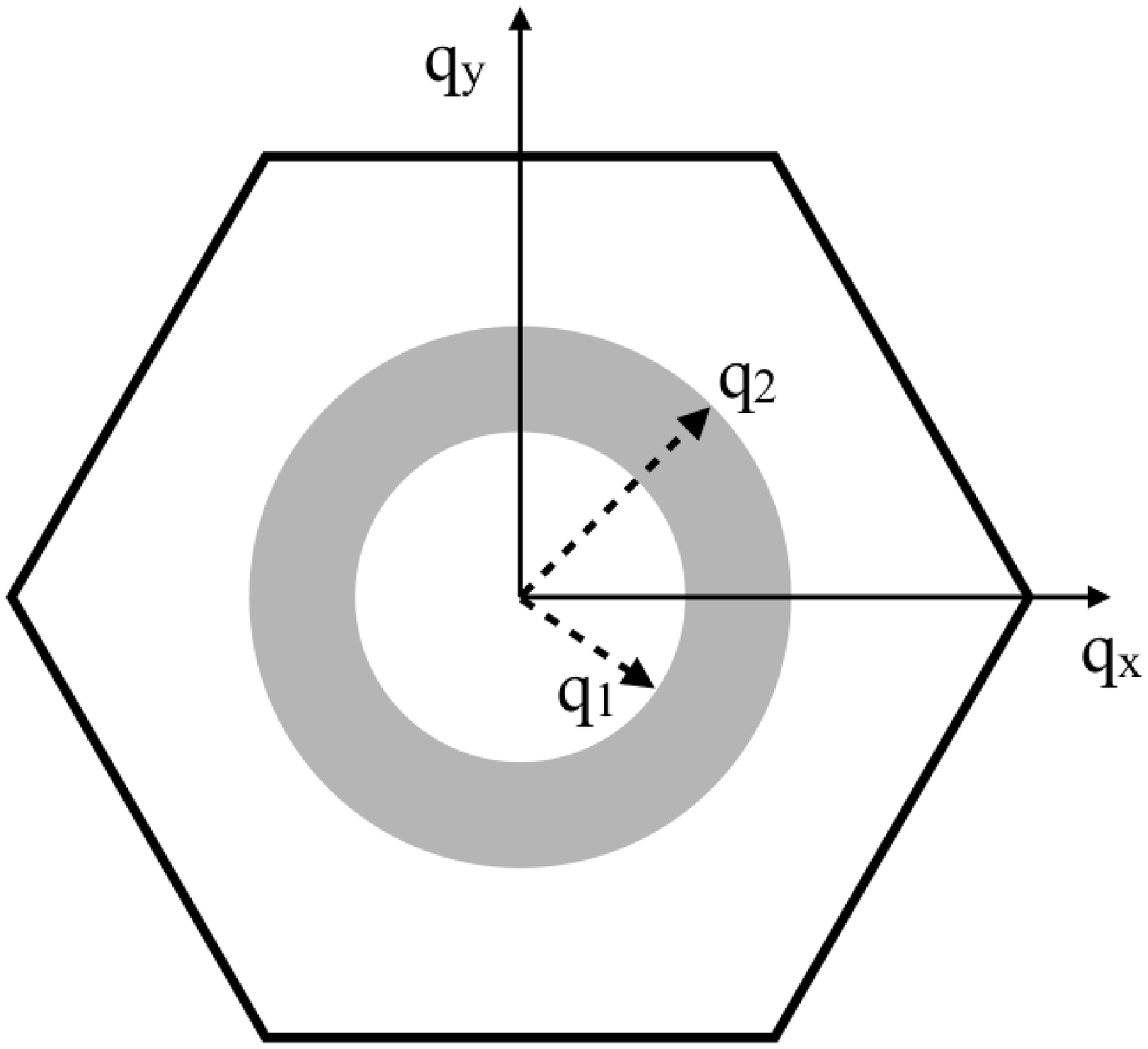}}
\subfloat[]
	{ \includegraphics[width=0.5\linewidth]{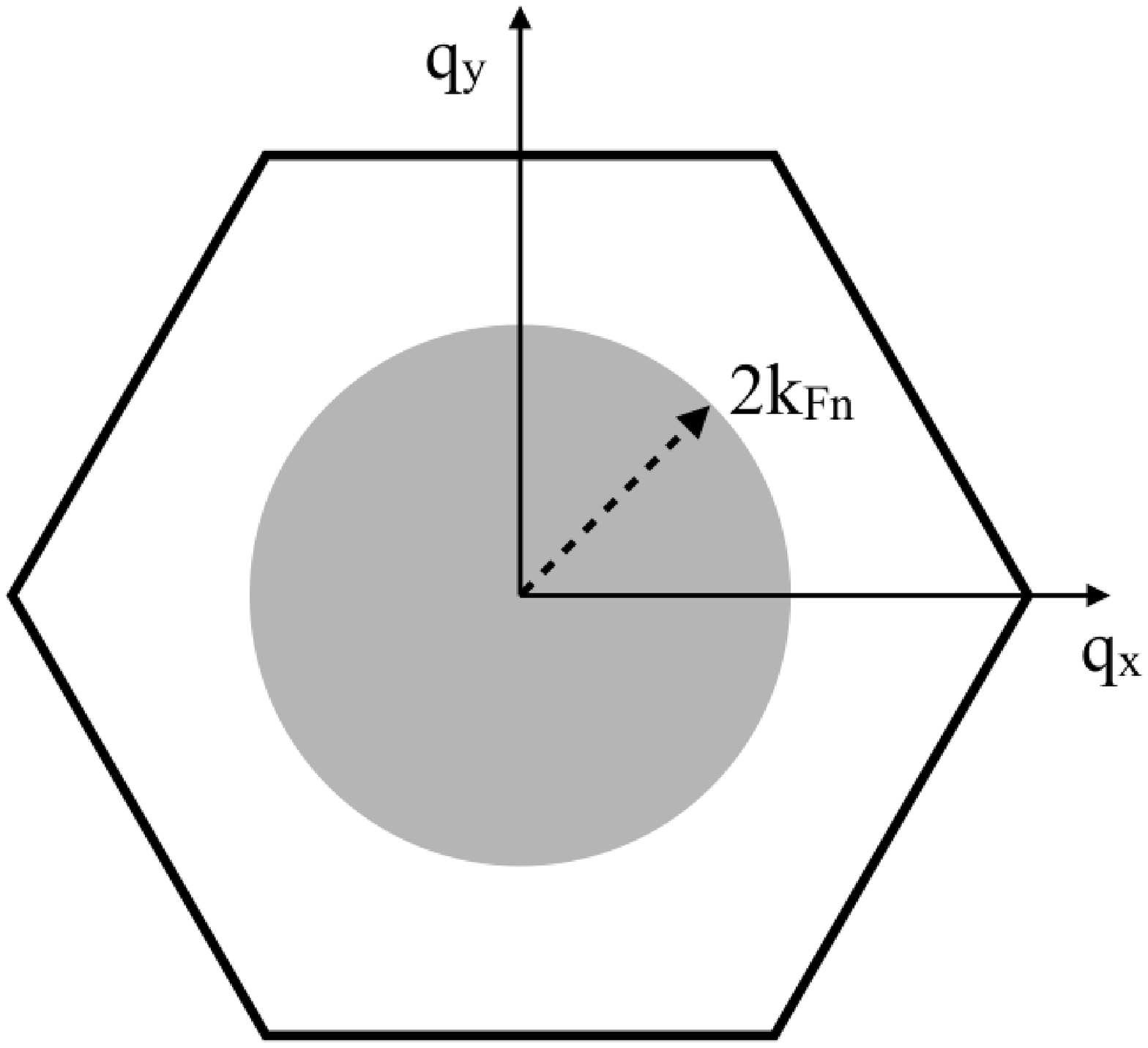}}\\
 \caption
   {Allowed regions of the Brillouin zone in $\vb{q}$-space (shaded regions) over which we consider the electron-phonon matrix elements to be constant in the case of  $n\ne m$ (a) and $n=m$  (b).}
   \label{fig:model_gamma}
   \vspace{-5mm}
\end{figure}

Consider now the case of distinct parabolic bands centered around the center of the First Brillouin Zone ($\Gamma$). The role of the Dirac deltas in Eq.~\ref{eq:spectral} is to limit scattering events from the $n$-th Fermi surface to the $m$-th one (i.e $\abs{\vb{k}}=k_{Fn}$ and $\abs{\vb{k}'}=k_{Fm}$, $\vb{k}'=\vb{k}+\vb{q}$). Thus the allowed values of $\abs{\vb{q}}$ fall in the shaded region of Fig.~\ref{fig:model_gamma}, i.e. in the range: 
 \begin{equation}
 \label{eq:q_squared}
 \begin{split}
 &q^2 = k^2+k'^2-2kk'\cos{\theta}=k_{Fn}^2+k_{Fm}^2-2k_{Fn}k_{Fm}\cos{\theta}\Rightarrow\\
 \Rightarrow&\abs{\vb{q}}\in\Biggl[  \sqrt{k_{Fn}^2+k_{Fm}^2-2k_{Fn}k_{Fm}}\text{ ; } \sqrt{k_{Fn}^2+k_{Fm}^2+2k_{Fn}k_{Fm}} \Biggr]\\
 &\in [q_1\text{;}q_2]
 \end{split}
 \end{equation}
Since it is an inexpensive computational task, we will compute electron-phonon matrix elements at $\vb{q}=\vb{\Gamma}$ and will consider them to be constant over the region delimited by Eq.~\ref{eq:q_squared}. As a consequence we will assume
\begin{equation}
\begin{split}
&\abs{g^\nu_{\vb{k}n,\vb{k}'m}}^2=\abs{g^\nu_{\vb{\Gamma}n,\vb{\Gamma}m}}^2, \hspace{1cm} \omega_{\vb{k'}-\vb{k},\nu}=\omega_{\vb{\Gamma},\nu}\\
&   \hspace{1cm} \text{for} \,\,\abs{\vb{k}}\in[0,k_{Fn}], \hspace{3mm}\abs{\vb{q}}\in[q_1,q_2]. \nonumber
\end{split}
\end{equation}

Using this simplifying assumption and inserting Eq.~\ref{eq:spectral} into Eqs.~\ref{eq:lambda} and \ref{eq:omega_log} gives, for a 2D electron gas:
\begin{subequations}
\begin{eqnarray}
 \label{eq:SSM2}
 \lambda = &&N_{\sigma}(0)\sum_{\nu}\frac{\langle g_{\nu\vb{\Gamma}}^2 \rangle}{\omega_{\vb{\Gamma}\nu}}   \label{eq:lambda_simple}\\
\log(\omega_{\textrm{log}})  = &&\frac{N_{\sigma}(0)}{\lambda}\sum_{\nu}\frac{\log(\omega_{\vb{\Gamma}\nu})\langle g_{\nu\vb{\Gamma}}^2 \rangle}{\omega_{\vb{\Gamma}\nu}}  \label{wlog3}\\
\alpha^2F(\omega)=&&\frac{N_{\sigma}(0)}{2}\sum_{\nu}\langle g_{\nu\vb{\Gamma}}^2 \rangle\delta(\omega-\omega_{\vb{\Gamma}\nu})\, ,\label{eq:a2F_simple}
\end{eqnarray}
\end{subequations}
where 
 \begin{equation}
 \label{eq:strength}
\langle g_{\nu\vb{\Gamma}}^2 \rangle =\sum_{n,m}\frac{\abs{g_{\vb{\Gamma} n,\vb{\Gamma} m}^\nu}^2N_{\sigma,n}(0)N_{\sigma,m}(0)}{N_{\sigma}^2(0)}
\end{equation}
is the squared average of the electron-phonon matrix elements over the Fermi surfaces, 
and $N_{\sigma,n}(0)$ is the total electronic density of states per spin at the Fermi level for the $n$-th Fermi surface.

\section{Results}
\subsection{Electronic structure}
As already observed in the hydrogenated diamond (111) surface  \cite{ref:RomaninAPSUSC2019}, hole doping by field effect induces an alteration of bond lengths even in H-Si (111). The affected atoms (Fig.~\ref{fig:atomic_structure}) are H(1) and the first three silicon atoms, Si(1), Si(2) and Si(3). Upon doping, the H-Si bond length has an increment of $\sim4\%$ and the angle $\theta_H$ increases by $\sim 1.6\%$. As for the silicon atoms, the Si(1)-Si(2) bond length increases by 0.8\% and the Si(2)-Si(3) one by 0.4 \%, while the other bonds are unperturbed (see Tab.\ref{tab:atomic_structure}).\\
\begin{table}[b]
  \begin{tabular}{ccc}
                            &  $n_{dop}=0\cdot10^{14}$ cm$^{-2}$  & $n_{dop}=6\cdot10^{14}$ cm$^{-2}$ \\
    \hline
     H-Si                & 1.51 \AA  & 1.57 \AA    \\
     Si(1)-Si(2)       & 2.36 \AA  & 2.38 \AA    \\
     Si(2)-Si(3)       & 2.36 \AA  & 2.37  \AA    \\
     Si(3)-Si(4)       & 2.37  \AA & 2.37   \AA    \\
     $\theta_H$      & $\ang{108.8}$ & $\ang{110.5}$ \\
     $\theta_{Si}$   & $\ang{110.1}$  & $\ang{108.5}$ \\
    \hline
  \end{tabular}
\caption{Atomic parameters in the undoped ($n_{dop}=0\cdot10^{14}$ cm$^{-2}$) and hole-doped ($n_{dop}=6\cdot10^{14}$ cm$^{-2}$) case for the hydrogenated silicon (111) surface, as labelled in Fig.~\ref{fig:atomic_structure}.} \label{tab:atomic_structure}
\end{table}

\begin{figure}
\centering
\includegraphics[width=0.9\linewidth]{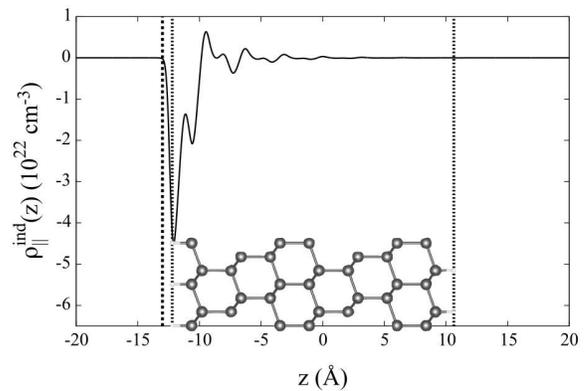}
\caption{Planar-averaged induced charge density ($\rho_{||}^{ind}(z)$) of the hole-doped hydrogenated silicon (111) surface with $n_{dop}=6\cdot10^{14}$ cm$^{-2}$.} \label{fig:charge_diff}
\end{figure}

We then study the spatial distribution of the induced holes in order to understand which layers are actually doped. In Fig.~\ref{fig:charge_diff} we plot the planar-averaged induced charge density $\rho_{||}^{ind}(z)$ along the $z$-axis:
\begin{equation}
\rho^{ind}_{||}(z)= \frac{1}{\Omega_{2D}}\int_{\Omega_{2D}}\Biggl\{\rho^{\text{h}}_{3D}(x,y,z)-\rho^{0}_{3D}(x,y,z)\Biggr\}dxdy
\end{equation}
where $\rho^{\text{h}}_{3D}$ ($\rho^{0}_{3D}$) is the 3D charge distribution in the hole-doped (undoped) case and $\Omega_{2D}$ is the unit cell surface area.
From Fig.~\ref{fig:charge_diff} we can clearly observe that the majority of the induced holes is concentrated on the hydrogen layer facing the metal gate and on the underlying $2-4$ carbon layers (i.e. a region of length $~5.57$ \AA). Therefore, field-effect doping is confined to the sample surface, in agreement with the field-induced atomic relaxation discussed above, and similarly to field-effect doped H-C(111) and MoS$_2$ \cite{ref:RomaninAPSUSC2019,ref:PiattiJPCM2019}, and does not affect atomic layers further into the slab, as can sometimes occur upon the application of electric fields of comparable magnitude \cite{ref:PiattiPRB2017, ref:UmmarinoPRB2017, ref:PiattiApSuSc2018}.\\

\begin{figure}
\centering
\includegraphics[width=0.8\linewidth]{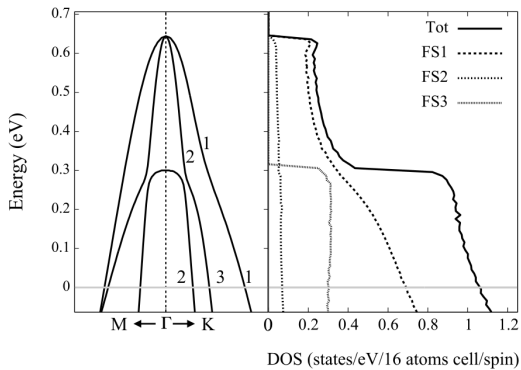}
\caption{Electronic structure anlaysis of the hydrogenated silicon (111) surface at $n_{dop}=6\cdot10^{14}$ cm$^{-2}$. First panel: electronic dispersion along high-symmetry directions of the First Brillouin Zone ($\vb{M}-\vb{\Gamma}-\vb{K}$). Second panel: total density of states (DOS) and the contribution of the three bands to the total DOS. The horizontal grey line is the Fermi level, here set to $E_F=0$. The DOS is in units of states/eV/16 atoms cell/spin. Labels 1 (FS1), 2 (FS2) and 3 (FS3) identify the bands crossed at the Fermi level.} \label{fig:el_str}
\end{figure}

\begin{figure}
\centering
\includegraphics[width=0.8\linewidth]{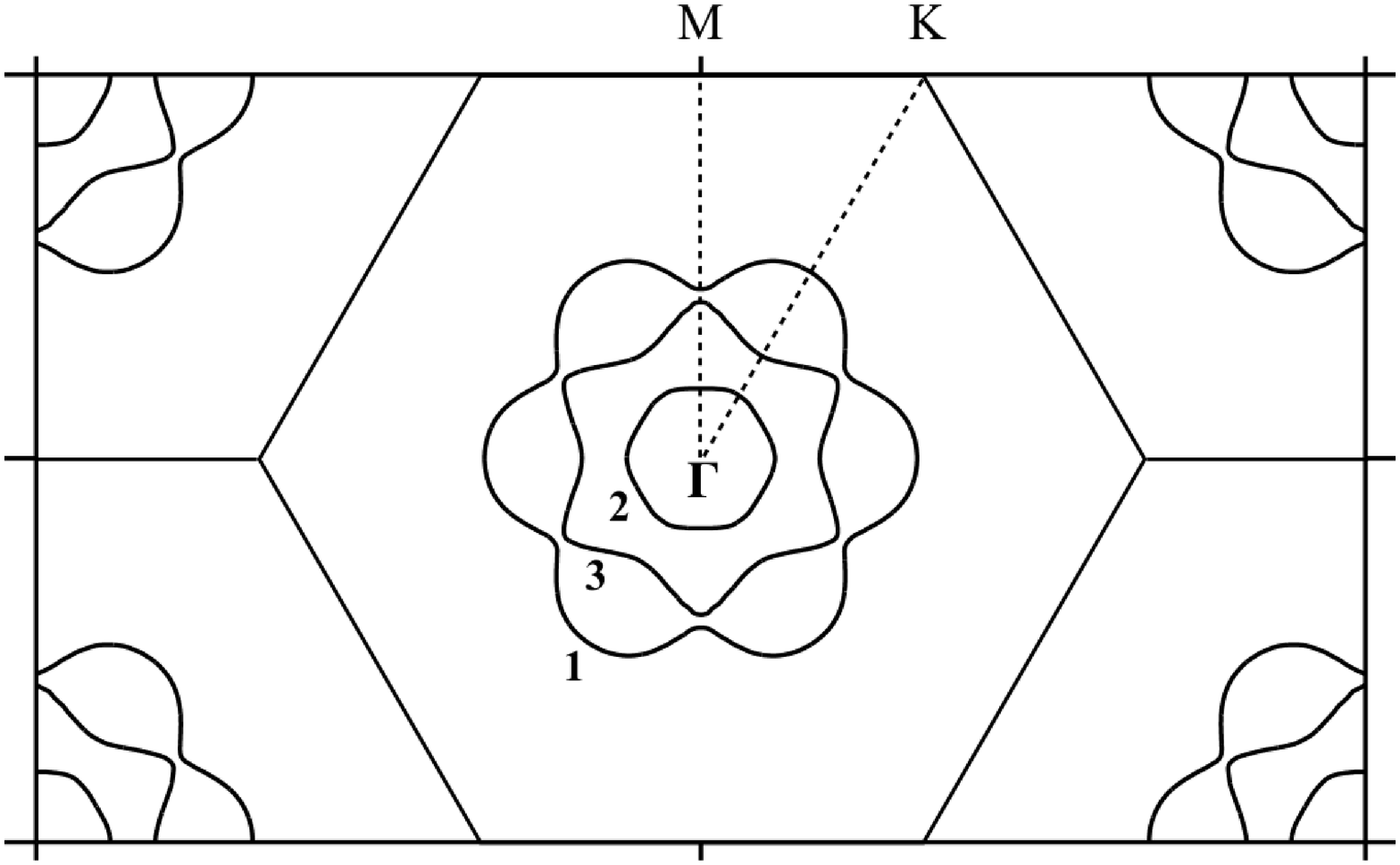}
\caption{2D Fermi surface of the hydrogenated silicon (111) system at $n_{dop}=6\cdot10^{14}$ cm$^{-2}$. $\vb{M}$, $\vb{\Gamma}$ and $\vb{K}$ are high symmetry points of the Brillouin zone. Labels 1,2,3 denote the different Fermi surface sheets in accordance with Fig.~\ref{fig:el_str}.} \label{fig:FS}
\end{figure}

We now focus our attention on the electronic band structure of the H-Si(111) surface, shown in the first panel of Fig.~\ref{fig:el_str}. Since we are inducing holes into the surface layers, the system is no longer insulating: a Fermi level appears ($E_F=0$ eV) crossing three valence bands centered around $\vb{k}=\vb{\Gamma}$ (the center of the Brillouin zone) which are almost parabolic. Moreover, around $0.3$ eV above the Fermi energy, there is an avoided crossing between the second and the third bands: therefore at the Fermi level they are swapped. In Fig.~\ref{fig:el_str} we keep track of this by labelling in the proper way the electronic bands. The distance between $E_F$ and the top of the first two valence bands is $\Delta E_{v,1}=\Delta E_{v,2}\approx640$ meV (since they are degenerate at $\vb{\Gamma}$), while the distance between $E_F$ and the top of the third valence bands is $\Delta E_{v,3}\approx296$ meV. As depicted in Fig.~\ref{fig:FS}, the Fermi surface is made up of three hole pockets centered around $\vb{k}=\vb{\Gamma}$.

The shape of the total electronic density of states (DOS) is step-like (Fig.~\ref{fig:el_str}, second panel), which is typical of 2D electron gases. The DOS at the Fermi level, $N_{\sigma}(0)$, is almost three times larger than in H-C(111) (see Tab.~\ref{tab:DOS}).  Moreover, the second band (FS2) does not contribute much to the total DOS with respect to the other two (FS1 and FS3), while in the case of H-C(111) FS2 and FS3 were equally relevant to the total DOS.
\begin{table}[b]
  \begin{tabular}{lcccc}
                            &  $N_{\sigma}(0)$  & $N_{\sigma,1}(0)$ & $N_{\sigma,2}(0)$ & $N_{\sigma,3}(0)$\\
    \hline
     H-C(111)  & 0.393  & 0.1783 & 0.1088 & 0.1059  \\
     H-Si(111)     & 1.040  & 0.6800 & 0.0700 &  0.2900   \\
    \hline
  \end{tabular}
\caption{Total DOS ($N_{\sigma}(0)$) and DOS per bands ($N_{\sigma,n}(0)$, $n=1,3$ as labelled in Fig.~\ref{fig:FS}) at the Fermi level ($E_F=0$) in units of states/eV/16 atoms cell/spin for the hole-doped hydrogenated diamond (111) \cite{ref:RomaninAPSUSC2019} and hydrogenated silicon (111) surfaces ($n_{dop}=6\cdot10^{14}$ cm$^{-2}$).} \label{tab:DOS}
\end{table}

\begin{figure}
\centering
\includegraphics[width=1.0\linewidth]{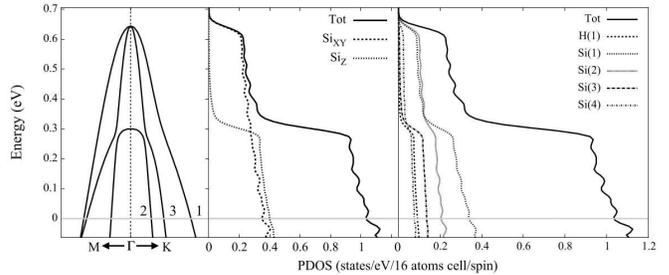}
\caption{Electronic structure anlaysis of the hydrogenated silicon (111) surface at $n_{dop}=6\cdot10^{14}$ cm$^{-2}$. First panel: electronic dispersion along high-symmetry directions of the First Brillouin Zone ($\vb{M}-\vb{\Gamma}-\vb{K}$). Second panel: partial density of states (PDOS) depicting the planar (XY) and out-of-plane (Z) character of the electronic bands. Third panel: partial density of states (PDOS) showing the contribution of the atomic layers involved in the field-effect doping (as labelled in Fig.~\ref{fig:atomic_structure}). The horizontal grey line is the Fermi level, here set to $E_F=0$. DOS and PDOS are in units of states/eV/16 atoms cell/spin.} \label{fig:partial_dos}
\end{figure}

We can also study the character of the electronic bands by looking at the partial density of states (PDOS). In the second panel of Fig.~\ref{fig:partial_dos} we show the in-plane (i.e. parallel to the surface, in the XY plane) and out-of-plane (i.e. perpendicular to the surface, along the Z axis) contributions to the total DOS. By comparison with the bandstructure reported in the first panel, it turns out that the first two bands (FS1 and FS2) are actually purely planar, while FS3 is a band involving out-of-plane orbitals. The third panel of Fig.~\ref{fig:partial_dos} reports the contribution to the DOS of the atomic layers affected by field-effect doping. Bands FS1 and FS2 are actually coming from the orbitals of Si(1) and Si(2), while Si(3), Si(4) and H(1) contribute to the third band (FS3). It is important to stress here the role of hydrogen atoms: while in the hydrogenated diamond (111) surface they did not contribute to the total DOS, in the H-Si(111) surface their contribution to the total DOS is comparable to that of Si(3) and Si(4).

\subsection{Superconductive properties}
\begin{figure}
\centering
\includegraphics[width=0.9\linewidth]{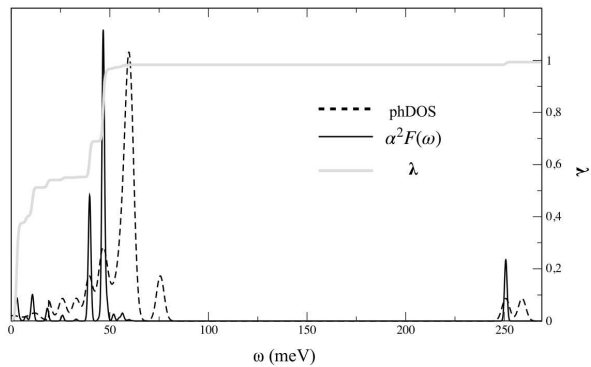}
\caption{Phonon density of states (phDOS, dashed line), Eliashberg spectral function ($\alpha^2F(\omega)$, solid black line) and electron-phonon coupling constant ($\lambda$, solid grey line) for the H-Si(111) surface at $n_{dop}=6\cdot10^{14}$ cm$^{-2}$, obtained from the simplified superconductive model. In order to plot the phDOS and $\alpha^2F(\omega)$ we replaced the $\delta(\omega-\omega_\nu)$ in Eq.~\ref{eq:a2F_simple} and Eq.~\ref{eq:ph_dos} with Gaussians of spread 0.002 eV and 0.001 eV respectively. The phDOS and $\alpha^2F(\omega)$ are in arbitrary units.} \label{fig:a2F}
\end{figure}

We finally turn our attention to the possible superconductive phase transition in the hole-doped hydrogenated silicon (111) surface by applying the simplified model described in Section \ref{sec:model}. In Fig.~\ref{fig:a2F} we plot the Eliashberg spectral function $\alpha^2F(\omega)$ (Eq.~\ref{eq:a2F_simple}), the electron-phonon coupling constant $\lambda$ (Eq.~\ref{eq:lambda_simple}) and the phonon density of states (phDOS) defined as:
\begin{equation}
\label{eq:ph_dos}
\text{phDOS}=\frac{1}{3N_{a}}\sum_{\nu}\delta(\omega-\omega_{\nu})
\end{equation}
where $N_{a}$ is the number of atoms in the unit cell. From the Eliashberg spectral function we can recognize four modes that have the strongest electron-phonon interactions. The first one is an out-of-plane vibrational mode at $\omega_{1,\perp}=39.80$ meV, with $\langle g_{1,\vb{\Gamma}}^2 \rangle=0.456\cdot10^{-2}$ eV$^2$, and is mainly due to H(1), Si(1) and Si(2). The second and third modes are degenerate at $\omega_{2-3,||}=46.64$ meV, whose strength is $\langle g_{2-3,\vb{\Gamma}}^2 \rangle =0.577\cdot10^{-2}$ eV$^2$, and describe an in-plane motion of H(1), Si(1) and Si(2). Finally we have the out-of-plane mode of H(1) at $\omega_{4,\perp}=250.79$ meV, with $\langle g_{4,\vb{\Gamma}}^2 \rangle =0.243\cdot10^{-2}$ eV$^2$. These values of the $\langle g_{\nu,\vb{\Gamma}}^2 \rangle$ are actually two orders of magnitude smaller than those we found in the hydrogenated diamond (111) surface, but this was to be expected since silicon atoms are heavier than carbon atoms and the electron-phonon matrix elements depend on the inverse of the atomic mass (Eq.~\ref{eq:elphon}).
\begin{table}[b]
	\begin{tabular}{lcccc}
		&  $\lambda$  & $\omega_{\textrm{log}}$ (cm$^{-1}$) & $\mu^*$ & $T_{\text{c}}$ (K)\\
		\hline
		H-C(111)  & 1.09  & 629.94 & $0.13-0.14$ & $63.14-57.20$  \\
		H-Si(111)     & 0.98  & 102.67 & $0.08-0.12$ &  $10.91-8.94$   \\
		\hline
		H-C(111) (W)  & 0.81  & 670.17 & $0.13-0.14$ & $34.93-29.60$  \\
		H-Si(111) (W)    & 0.73  & 109.22 & $0.08-0.12$ &  $6.98-5.14$   \\
		\hline
	\end{tabular}
	\caption{Electron-phonon coupling constant ($\lambda$), logarithmic averaged phonon frequency ($\omega_{\textrm{log}}$), Morel-Anderson pseudopotential ($\mu^*$) and superconductive critical temperature ($T_{\text{c}}$) for hole-doped H-C(111) \cite{ref:RomaninAPSUSC2019} and H-Si(111) surfaces ($n_{dop}=6\cdot10^{14}$ cm$^{-2}$). The last two lines report the same quantities for the Wannier-renormalized systems.} \label{tab:super}
\end{table}

Nevertheless, the electron-phonon coupling constant $\lambda$ computed with the simplified superconductive model for the H-Si(111) surface is comparable to that obtained in the same way for H-C(111), due to the higher densities of states at the Fermi level of the former which compensate for the lower electron-phonon matrix elements: As we can see from Tab~\ref{tab:super}, in the H-Si(111) surface we have $\lambda_{Si}=0.98$ while for H-C(111) we have $\lambda_{C}=1.09$. However the $T_{\text{c}}$ of the hole-doped H-Si(111) surface is $\sim6$ times smaller than that H-C(111): indeed we find $T_{\text{c}}\in[8.94;10.91]$ K, with $\mu^*\in[0.08;0.12]$. This is due to the fact that hole-doped H-C(111) has a logarithmic averaged phonon frequency $\omega_{\textrm{log}}$ which is $\sim6$ times bigger than that of H-Si(111). 
As a final remark, we have to recall that in H-C(111) this simplified superconductive model was found to overestimate $\lambda$ and underestimate $\omega_{\text{log}}$, as discussed in Ref.\cite{ref:RomaninAPSUSC2019}. Indeed, a more accurate calculation of the electron-phonon matrix elements through a Wannier-interpolation scheme over the whole Brillouin zone \cite{ref:MostofiCPC2014,ref:CalandraPRB2010} gave the values of $\lambda$ and $\omega_{\textrm{log}}$ reported in the third line of Tab.\ref{tab:super}, which are renormalized by $\sim30\%$ and $\sim3\%$ respectively. If similar renormalizations were to occur also for H-Si(111), we would obtain the values reported in the fourth line of Tab.\ref{tab:super}, i.e. $\omega_{\textrm{log}}=109.22 \mathrm{cm^{-1}}$ and $\lambda=0.73$, which still give a superconductive transition with a reduced critical temperature $T_{\text{c}}\in[5.14;6.98]$ K, depending on the value of $\mu^*$.

\section{Conclusions}

In this work we have shown, by means of DFT calculations \cite{ref:QE1,ref:QE2,ref:SohierPRB2017}, that the hydrogenated (111) surface of silicon can develop a superconducting phase if sufficiently hole-doped by field effect - in particular, at the surface hole concentration $n_{dop}=6\cdot10^{14}$ cm$^{-2}$. We have estimated the superconductive critical temperature $T_{\text{c}}$ via the McMillan/Allen-Dynes formula and a simplified superconductive model based on linear response computations at $\vb{q}=\vb{\Gamma}$. We have compared the results with the field-effect hole-doped hydrogenated diamond (111) surface at the same doping value \cite{ref:RomaninAPSUSC2019}. Due to the smaller atomic mass, the H-Si(111) surface shows lower values of the electron-phonon matrix elements with respect to the diamond counterpart. However, thanks to the larger density of states at the Fermi level, we have found a sizable $\lambda_{Si}=0.98$ (while that of H-C(111) was $\lambda_{C}=1.09$). Nevertheless, the low $\omega_{\textrm{log}}$ of the H-Si(111) surface allows for a superconductive phase transition at $T_{\text{c}}\in[8.94;10.91]$ K (with $\mu^*\in[0.08;0.12]$, Refs.~ \cite{ref:JinJPCM1997, ref:BourgeoisAPL2007}), which is $\sim6$ times smaller than the diamond counterpart. Note that this value is an upper limit for $T_{\text{c}}$, because the simplified model was shown \cite{ref:RomaninAPSUSC2019} to overestimate it; A more accurate evaluation could be obtained through a Wannier interpolation of the electron-phonon matrix elements over the whole Brillouin zone.
As we already observed for the hydrogenated diamond surface, the superconducting phase is spatially limited to the first few layers of the sample: indeed, the analysis of the distribution of the induced charge and of the electronic and vibrational properties shows that H(1), Si(1) and Si(2) atoms are the ones contributing the most to the superconductive phase.

\acknowledgments
The author acknowledges Th. Sohier, M. Calandra, D. Daghero and E. Piatti for fruitful discussions. Computational resources were provided by hpc@polito (http://hpc.polito.it).


\begin{thebibliography}{0}

\bibitem{ref:CohenPR1964}  {Cohen~M.L.}  \textit{Phys. Rev.} \textbf{134},  {A511} (1964).
\bibitem{ref:ConnetablePRL2003}  {Connetable~D., Timoshevskii~V., Masenelli~B., Beille~J., Marcus~J., Barbara~B., Saitta~A.M., Rignanese~G.M., Melinon~P., Yamanaka~S. and Blase~X.}  \textit{Phys. Rev. Lett.} \textbf{91}, {247001} (2003).
\bibitem{ref:BoeriPRL2007} {Boeri~L., Kortus~J. and Andersen~O.K.}  \textit{Phys. Rev. Lett.} \textbf{93}, {237002} (2004).
\bibitem{ref:BlaseNM2009} {Blase~X., Bustarret~E., Chapelier~C., Klein~T. and Marcenat~C.}  \textit{Nature Materials} \textbf{8}, {375-382} (2009).
\bibitem{ref:EkimovNat2004} {Ekimov~E.A., Sidorov~V.A., Bauer~E.D., Mel'nik~N.N, Curro~N.K., Thompson~J.D. and Stishov~S.M.}  \textit{Nature} \textbf{428}, {542-545} (2004).
\bibitem{ref:BustarretNat2006} {Bustarret~E., Marcenat~C., Achatz~P., Kacmarcik~J., Levy~F., Huxley~A., Ortega~L., Bourgeois~E., Blase~X., Debarre~D. and Boulmer~J.}  \textit{Nature} \textbf{444}, {465-468} (2006).
\bibitem{ref:DagheroPRL2012} {Daghero~D., Paolucci~F., Sola~A., Tortello~M., Ummarino~G.A., Agosto~M., Gonnelli~R.S., Nair~J.R. and Gerbaldi~C.}  \textit{Phys. Rev. Lett.} \textbf{108}, {066807} (2012).
\bibitem{ref:PiattiPRB2017}  {Piatti~E., Daghero~D., Ummarino~G.A., Laviano~F., Nair~J.R., Cristiano~R., Casaburi~A., Portesi~C., Sola~A. and Gonnelli~R.S.}  \textit{Phys. Rev. B} \textbf{95}, {140501} (2017).
\bibitem{ref:PiattiLTP2019} {Piatti~E., Romanin~D., Daghero~D. and Gonnelli~R.S.}  \textit{Low Temp. Phys.} \textbf{45(11)}, {1143-1155} (2019).
\bibitem{ref:NakamuraPRB2013} {Nakamura~K., Rhim~S.H., Sugiyama~A., Sano~K., Akiyama~T., Ito~T., Weinert~M and Freeman~A.J.}  \textit{Phys. Rev. B} \textbf{87}, {214506} (2013).
\bibitem{ref:SanoPRB2017} {Sano~K., Hattori~T. and Nakamura~K.}  \textit{Phys. Rev. B} \textbf{96}, {155144} (2017) .
\bibitem{ref:RomaninAPSUSC2019}  {Romanin~D., Sohier~Th., Daghero~D., Mauri~F., Gonnelli~R. S.  and Calandra~M.}  \textit{Appl. Surf. Sci.} \textbf{496}, {143709} (2019).
%
\bibitem{ref:Wyckoff1963}  {Wyckoff~R.W.G.} in \textit{Crystal Structures, 2nd ed}, (Interscience Publishers, New York) 1963, Vol. 1, pp~7-83.
\bibitem{ref:QE1}   {Giannozzi~P. et al.}  \textit{J. Phys.: Condens. Matter} \textbf{21}, {395502}  (2009).
\bibitem{ref:QE2}   {Giannozzi~P. et al.}  \textit{J. Phys.: Condens. Matter} \textbf{29}, {465901} (2017).
\bibitem{ref:SohierPRB2017}  {Sohier~Th., Calandra~M. and Mauri~F.}  \textit{Phys. Rev. B} \textbf{96}, {075448} (2017).
\bibitem{ref:UFP} {Prandini~G., Marrazzo~A., Castelli~I.E., Mounets~N. and Marzari~N.} in  \textit{A Standard Solid State Pseudopotentials (SSSP) library optimized for precision and efficiency (Version 1.1, data download), Materials Cloud Archive} 2018, doi: 10.24435/materialscloud:2018.0001/v3.
%
 \bibitem{ref:McMillanPR1968}  {McMillan~W. L.}  \textit{Phys. Rev.} \textbf{167}, {331} (1968).
\bibitem{ref:AllenPRB1975}  {Allen~P. B. and Dynes~R. C.}  \textit{Phys. Rev. B} \textbf{12}, {905} (1975).
\bibitem{ref:JinJPCM1997} {Jin~Y.G., Lee~K.-H. and Chang~K. J.}  \textit{J. Phys.: Condens. Matter} \textbf{9}, {6351-6358} (1997).
\bibitem{ref:BourgeoisAPL2007} {Bourgeois~E. and Blase~X.}  \textit{Appl. Phys. Lett.} \textbf{90}, {142511} (2007).
\bibitem{ref:PiattiJPCM2019}  {Piatti~E., Romanin~D. and Gonnelli~R.S.}  \textit{J. Phys. Condens Matter} \textbf{31}, {114002} (2019).
\bibitem{ref:UmmarinoPRB2017}  {Ummarino~G.A., Piatti~E., Daghero~D., Gonnelli~R.S., Sklyadneva~I.Y., Chulkov~E.V. and Heid~R.}  \textit{Phys. Rev. B} \textbf{96}, {064509} (2017).
\bibitem{ref:PiattiApSuSc2018}  {Piatti~E., Romanin~D., Gonnelli~R.S. and Daghero~D} \textit{Appl. Surf. Sci.} \textbf{461}, {17-22} (2018).
%
\bibitem{ref:MostofiCPC2014} {Mostofi~A.A., Yates~J.R., Pizzi~G., Lee~Y.S., Souza~I., Vanderbilt~D. and Marzari~N.}  \textit{Comput. Phys. Commun.} \textbf{185}, {2309} (2014).
\bibitem{ref:CalandraPRB2010} {Calandra~M., Profeta~G. and Mauri~F.}  \textit{Phys. Rev. B} \textbf{82}, {165111} (2010).
%

\end{thebibliography}
\end{document}